\documentclass[sigconf]{acmart}

\usepackage{amsmath, bm}
\usepackage{multirow}
\usepackage{multicol}
\usepackage{subcaption}
\usepackage{hhline}

\usepackage{float}
\usepackage{placeins}
\usepackage{varioref}
\usepackage{bbm}
\usepackage{dsfont}
\usepackage{enumitem}
\usepackage{algorithm}
\usepackage[noend]{algpseudocode}
\algnewcommand{\LineComment}[1]{\State \(\triangleright\) #1}   

\newtheorem{remark}{Remark}

\renewcommand\footnotetextcopyrightpermission[1]{}
\AtBeginDocument{%
  \providecommand\BibTeX{{%
    \normalfont B\kern-0.5em{\scshape i\kern-0.25em b}\kern-0.8em\TeX}}}

\begin{document}


\title{Contrastive Counterfactual Learning for Causality-aware Interpretable Recommender Systems}
\author{Guanglin Zhou}
\affiliation{%
  \institution{The University of New South Wales}
  \city{Sydney}
  \country{Australia}
  \postcode{2052}
}
\orcid{0000-0002-1305-2622}
\email{guanglin.zhou@unsw.edu.au}

\author{Chengkai Huang}
\affiliation{%
  \institution{The University of New South Wales}
  \city{Sydney}
  \country{Australia}
  \postcode{2052}
}
\orcid{0000-0002-1630-424X}
\email{chengkai.huang1@unsw.edu.au}

\author{Xiaocong Chen}
\affiliation{%
  \institution{The University of New South Wales}
  \city{Sydney}
  \country{Australia}
  \postcode{2052}
}
\orcid{0000-0002-8849-4943}
\email{xiaocong.chen@unsw.edu.au}

\author{Xiwei Xu}
\affiliation{%
  \institution{Data61, CSIRO}
  \city{Eveleigh}
  \country{Australia}
  \postcode{2052}
}
\orcid{0000-0002-2273-1862}
\email{xiwei.xu@data61.csiro.au}

\author{Chen Wang}
\affiliation{%
  \institution{Data61, CSIRO}
  \city{Eveleigh}
  \country{Australia}
  \postcode{2052}
}
\orcid{0000-0002-3119-4763}
\email{chen.wang@data61.csiro.au}

\author{Liming Zhu}
\affiliation{%
  \institution{Data61, CSIRO}
  \city{Eveleigh}
  \country{Australia}
  \postcode{2052}
}
\orcid{0000-0001-5839-3765}
\email{liming.zhu@data61.csiro.au}

\author{Lina Yao}
\affiliation{%
  \institution{Data61, CSIRO}
  \city{Eveleigh}
  \country{Australia}
}
\affiliation{%
  \institution{The University of New South Wales}
  \city{Sydney}
  \country{Australia}
  \postcode{2052}
}
\orcid{0000-0002-4149-839X}
\email{lina.yao@unsw.edu.au}
  
    

\renewcommand{\shortauthors}{Guanglin Zhou et al.}

\begin{abstract}

The field of generating recommendations within the framework of causal inference has seen a recent surge, with recommendations being likened to treatments. 
This approach enhances insights into the influence of recommendations on user behavior and helps in identifying the underlying factors. 
Existing research has often leveraged propensity scores to mitigate bias, albeit at the risk of introducing additional variance. 
Others have explored the use of unbiased data from randomized controlled trials, although this comes with assumptions that may prove challenging in practice. 
In this paper, we first present the causality-aware interpretation of recommendations and reveal how the underlying exposure mechanism can bias the maximum likelihood estimation (MLE) of observational feedback. 
Recognizing that confounders may be elusive, we propose a contrastive self-supervised learning to minimize exposure bias, employing inverse propensity scores and expanding the positive sample set. 
Building on this foundation, we present a novel contrastive counterfactual learning method (CCL) that incorporates three unique positive sampling strategies grounded in estimated exposure probability or random counterfactual samples. 
Through extensive experiments on two real-world datasets, we demonstrate that our CCL outperforms the state-of-the-art methods.

\end{abstract}



\maketitle

\section{Introduction}
\begin{figure}[!hb]
\centering
\includegraphics[width=0.99\linewidth]{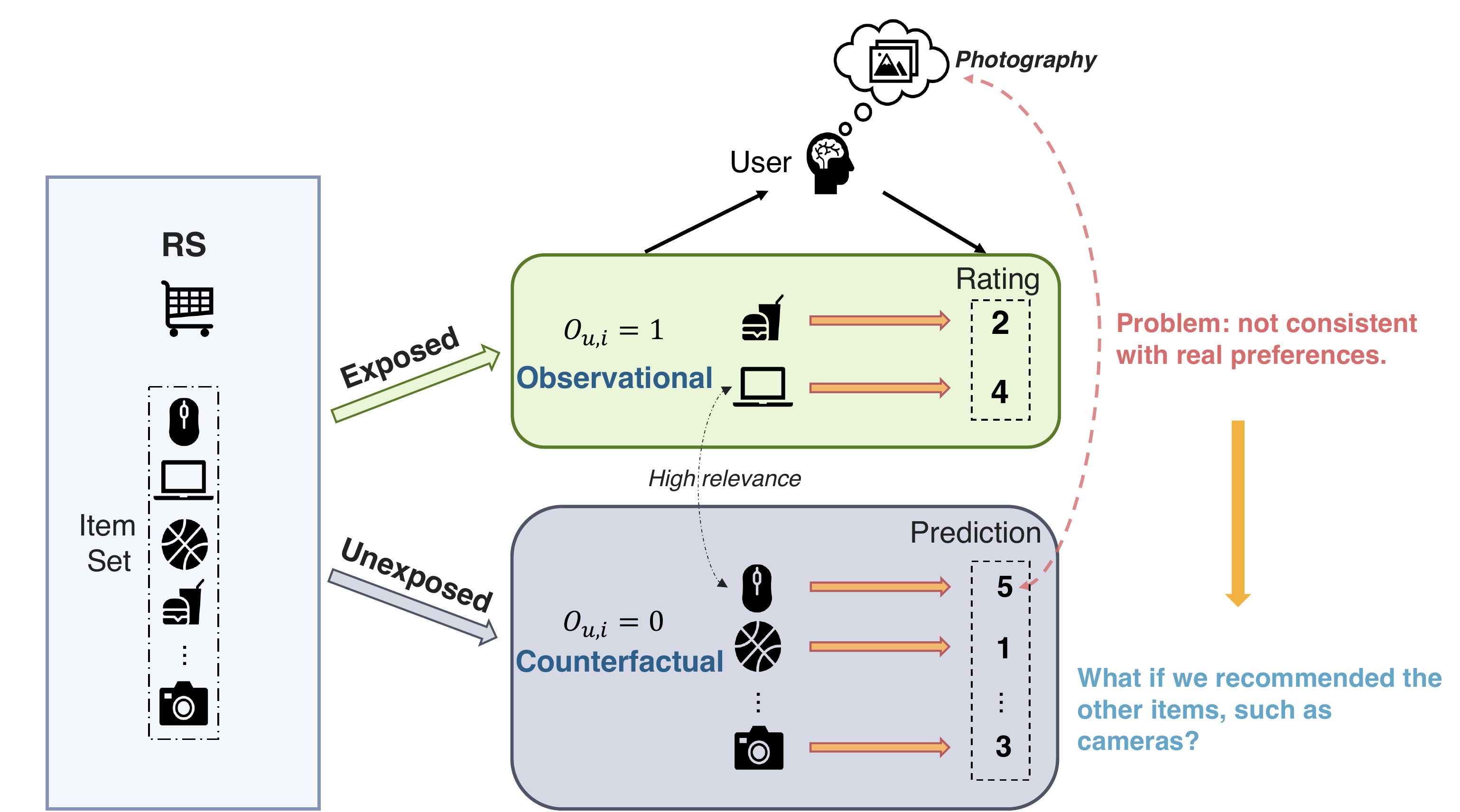}

\caption{A motivation illustration of formulating recommendations in the context of causal inference. 
}
\label{266492058979}
\end{figure}

Recommender systems are commonly used in a range of applications, including search engines, ad placement, and e-commerce websites, which generate feedback data based on two factors: exposures and ratings. 
Chronologically, the system initially presents users with a list of recommended items, which they can then click and rate based on their likes or dislikes. 
Exposures refer to the fact that users may not able to observe all items in each interaction, and the recommender system determines the mechanism for exposing items to users.
Ratings, on the other hand, express explicit preferences and reflect users' interests in items. 
It is important to recognize that we can only reveal users' preferences based on the items they are exposed to. 
If a user does not see an item, we cannot determine their opinion of it, which can lead to biases in our understanding of their preferences. 

In recent years, research has increasingly employed causal inference frameworks for recommendations \cite{schnabel2016recommendations,liang2016causal,wang2020causal,gao2022causal}. Recommendations are viewed as treatments, and traditional methods like matrix factorization use observational data to predict user preferences, as illustrated in Figure \ref{266492058979}. However, these approaches can be biased by the exposure mechanism, overshadowing true preferences. Accurate preferences can be inferred when the user has viewed each item or when recommendations are random, aligning with causal inference. This perspective corrects exposure's influence, enhancing prediction accuracy.

Research on recommendation from the perspective of causal inference has evolved into three main categories: (1) Propensity score-based methods \cite{schnabel2016recommendations, swaminathan2015self} integrate inverse propensity scores into standard models but face challenges in variance and stability. Extensions \cite{jiang2016doubly, wang2019doubly, li2022stabilized} provide more robust estimators. (2) Side information-based techniques \cite{bonner2018causal, wang2021combating, li2021causal, sheth2022causal} utilize unbiased data or network information to re-weight biased ratings and mitigate biases, though obtaining auxiliary knowledge may be limiting. (3) Information bottleneck-based methods \cite{wang2020information, liu2021mitigating} focus on learning unbiased representations but are less effective in understanding the data generating process and reducing exposure bias.

\begin{figure*}[!htb]
    \centering
    \includegraphics[width=0.99\linewidth, height=0.52\textwidth, keepaspectratio]{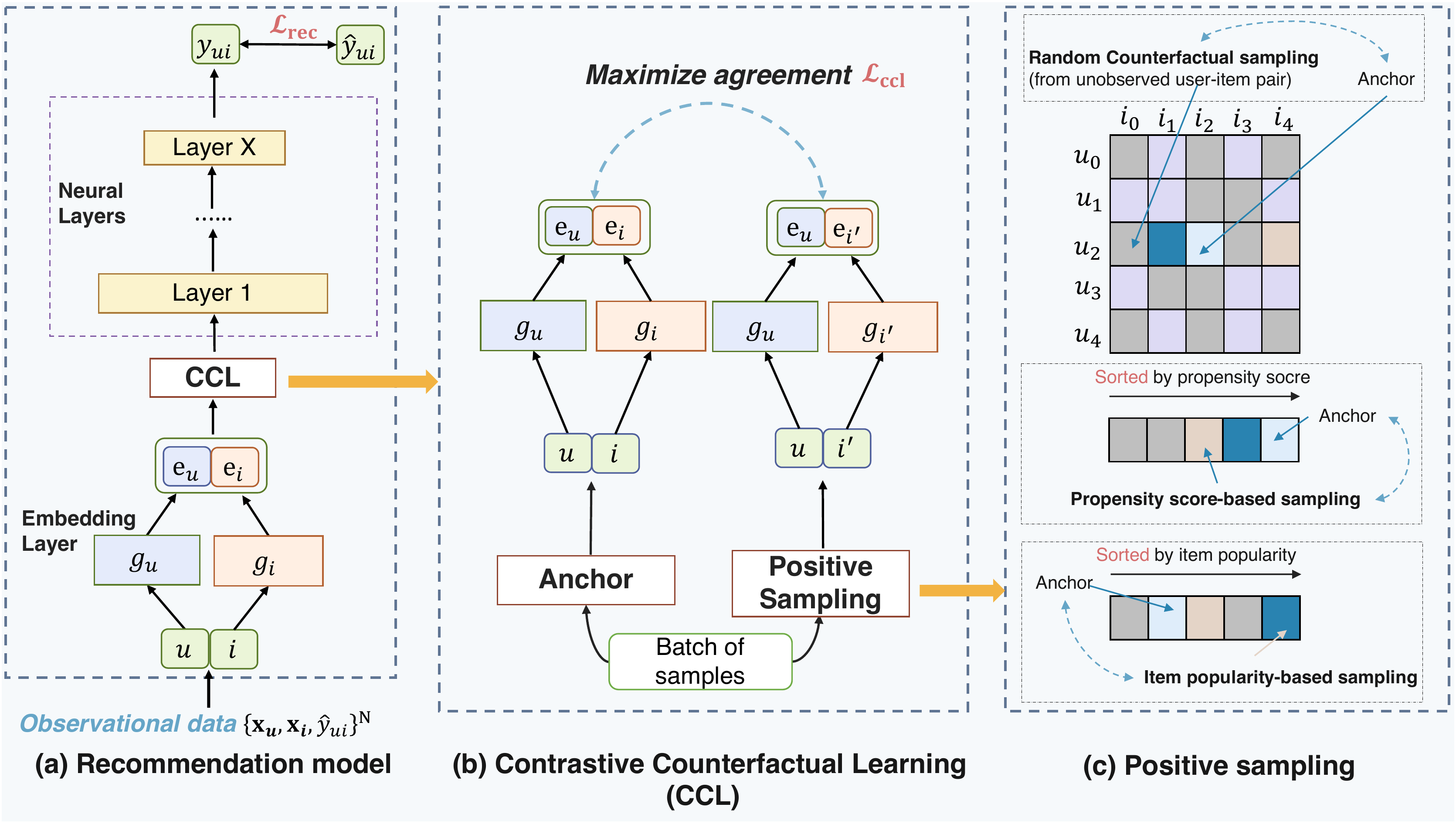}
    \caption{Overall framework. 
    (a) the structure of the recommendation model that incorporates a novel CCL module, where $\mathcal{L}_{rec}$ denotes the empirical risk for preference prediction.
    (b) the CCL module itself, which employs three novel sampling methods, with the goal of maximizing the agreement within two views through the use of $\mathcal{L}_{ccl}$. 
    (c) three novel sampling methods, including random counterfactual sampling, propensity score-based, and item popularity-based sampling.}
    \label{927512613774}
\end{figure*}

Despite the successes of existing models, it is necessary to model the data generating process in order to more fully understand the causality behind the data and identify potential confounders.
Structured causal models (SCMs) offer potential assistance, with recent studies proposing causal graphs for representing causal generation processes \cite{zhang2021causal, wang2021deconfounded, wei2021model}. 
However, these may fall short in generalizing to various biases and effectively modeling the exposure mechanism. 
To tackle exposure bias, we propose a strategy targeting underlying exposure mechanisms, employing contrastive self-supervised learning (SSL), a potent technique for mitigating such bias.
In this research, we employ structural causal models (SCMs) to introduce a causal graph representing the data generating process in recommender systems, identifying confounders and emphasizing the need to address exposure bias for accurate preference modeling. 
We propose contrastive self-supervised learning (SSL) with inverse propensity scores and introduce a novel method, contrastive counterfactual learning (CCL), combining three positive sampling strategies.

This work makes several important contributions to the field:
\begin{itemize}
    \item We highlight the importance of considering causal interpretation in recommendations and demonstrate the need to mitigate the exposure bias. 
    To address this issue, we propose the use of contrastive SSL, specifically through the implementation of inverse propensity scores and the expansion and diversification of the positive sample set. 
    \item We propose an innovative contrastive counterfactual learning method (CCL) that incorporates three novel positive sampling strategies based on estimated exposure probability or random counterfactual samples, which are found to be both effective and informative. 
    \item Our research reveals that sampling from random counterfactual sets leads to superior performance, highlighting the value of a large number of counterfactual samples in data. 
    Through extensive experimental results on two real-world benchmark datasets, we demonstrate the effectiveness of our method in comparison to ten baseline methods. 
    
\end{itemize}

\section{Preliminary}
\label{sec_problem}

In this study, we consider two sets of users and items, denoted as $\mathcal{U}$ and $\mathcal{I}$, respectively.
The number of elements in $\mathcal{U}$ and $\mathcal{I}$ are represented by $m$ and $n$, respectively, with each element denoted as $u \in \{1,\cdots,m\}$ and $i \in \{1,\cdots,n\}$.

The observational feedback data, denoted as $\mathcal{D}$, contains two types of data chronologically: exposures and ratings.
We use the indicator $O_{u,i} \in \{0,1\}$ to represent the exposure status, and $\hat{y}_{u,i}$ to represent the true rating for a given user-item pair. 
It is worth noting that if $O_{u,i}=0$, the value of $\hat{y}_{u,i}=0$ is missing and unknown.
In some cases, an item may be recommended to a user ($O_{u,i}=1$), but the user may not like it ($\hat{y}_{u,i}=0$).
On the other hand, a user may see some other items ($O_{u,i}=1$) with preferences ($\hat{y}_{u,i}=1$).

Formally, standard recommendation models are based on maximum likelihood estimation:
\begin{equation}
    \label{551497509538}
    \mathcal{L}_{MLE} = \frac{1}{\vert \{ (u,i): O_{u,i}=1\} \vert} \sum_{(u,i):O_{u,i}=1}\delta(\hat{y}_{u,i}, y_{u,i})
\end{equation}
where $\delta$ serves as the chosen loss function. 
However, it should be noted that $\mathcal{L}_{MLE}$ is not an unbiased estimation of true preferences due to the impact of the underlying exposure mechanism:
\begin{equation}
    \label{551497509539}
    \mathbb{E}[\mathcal{L}_{MLE}] \neq \mathcal{L}: \frac{1}{\vert \mathcal{U} \vert \cdot \vert \mathcal{I} \vert} \sum_{u=1}^{m} \sum_{i=1}^{n}\delta(\hat{y}_{u,i}, y_{u,i})
\end{equation}

The causal view in recommendations posits that the underlying exposure mechanism, which may be biased (impacted by popular items), determines what each user sees and impacts the inference of their preferences \cite{schnabel2016recommendations, liang2016causal}. 
As depicted in Figure \ref{266492058979}, the recommendation model may mistakenly assume that the user prefers a mouse and computer, while in reality the user is interested in photography. 
This naturally prompts the question of "What would happen if the user was recommended the camera?", which can be addressed through regarding the rating prediction as solving for treatment effects in causal inference, as represented by $y_{u,i}$ when $O_{u,i}=0$ is replaced with $O_{u,i}=1$ \cite{wang2020causal}.
In order to more accurately simulate experimental settings, some researchers have utilized inverse propensity scores (IPS) to re-weight observational data. 
This method allows for the data to be manipulated as if the items were randomly assigned to each participant, rather than being determined by the underlying exposure mechanism:

\begin{equation}
\label{653054168312}
    {\mathcal{L}}_{IPS} =  \frac{1}{\vert \{ (u,i): O_{u,i}=1\} \vert} \sum_{(u,i):O_{u,i}=1} \frac{\delta(\hat{y}_{u,i}, y_{u,i})}{P_{u,i}} 
\end{equation}

The propensity score, represented as $P_{u,i}$, is regarded as the probability of exposure for a given user-item pair.

In this study, we aim to investigate the use of contrastive SSL techniques for manipulating representations in order to mitigate the limitations of previous works, such as extra variances introduced by IPS-based methods. 

More formally, as shown in Figure \ref{927512613774}, we represent raw feature vector for the user $u$ as $\mathbf{x}_u$, which may include user id, gender, age, etc., and the raw feature vector for the item $i$, $\mathbf{x}_i$, separately. 

These vectors are encoded using encoders $g_u$ and $g_i$, respectively, yielding representations $\mathbf{h}_u$ and $\mathbf{h}_i$. 
We then concatenate these representations to create $\mathbf{h}_{u,i}$, which serves as the input for a stack of neural network layers $f(\cdot)$, to predict the rating $y_{u,i}=f(\mathbf{h}_{u, i})$.  

The recommendation loss is calculated using $\delta(\hat{y}_{u,i},y_{u,i}))$, where the term $\delta$ refers to the loss function, i.e.,  log-loss:    
\begin{equation}
\label{l_rec}
\begin{aligned}
     l_{rec}(u,i)  &= -\hat{y}_{u,i}\cdot log(y_{u,i}) - (1-\hat{y}_{u,i})\cdot log(1-y_{u,i}) \\
     &= -\hat{y}_{u,i}\cdot log(f(\mathbf{h}_{u,i})) - (1-\hat{y}_{u,i})\cdot log(1-f(\mathbf{h}_{u,i}))\\
\end{aligned}
\end{equation}
Our goal is to manipulate the representations $\mathbf{h}_{u,i}$ so that they resemble samples from an experimental distribution where items are exposed randomly to each user.
By doing so, we aim to provide more accurate predictions of users' natural preferences with $f(\cdot)$. 
In the next section, we will discuss the methods used to distort representations $\mathbf{h}_{u,i}$ and model the prediction function $f(\cdot)$.

\section{The Proposed Method}
\label{sec_method}

\subsection{Causality-Aware Interpretation of Recommendations}
\label{sec_IREC}

In this section, we argue that the utilization of maximum likelihood estimation (MLE) (as represented in Eq.\ref{551497509539}), which focuses on $P(Y\vert U, I)$, may lead to spurious effects stemming from the underlying exposure mechanism.
Specifically, this approach may lead us to mistakenly assume that the user prefers the mouse over the camera, as depicted in Figure \ref{266492058979}. 
One possible solution to this issue is the use of structural causal models (SCMs) for data generation \cite{Pearl2000CausalityMR, Bareinboim2022OnPH}. 
Recent works \cite{wang2021deconfounded, wei2021model} also illustrate the potential for SCMs to identify the popularity bias behind data. 
To this end, we introduce a causal graph to represent the typical variables in the recommendation models. 
In contrast to the association $P(Y\vert U,I)$, we propose blocking the confounder path, which is equivalent to an imagined experimental scenario where each item is exposed with equal probability, as proposed by \cite{liang2016causal}. 
 
\begin{figure}[!ht]
\centering

\includegraphics[width=0.8\linewidth]{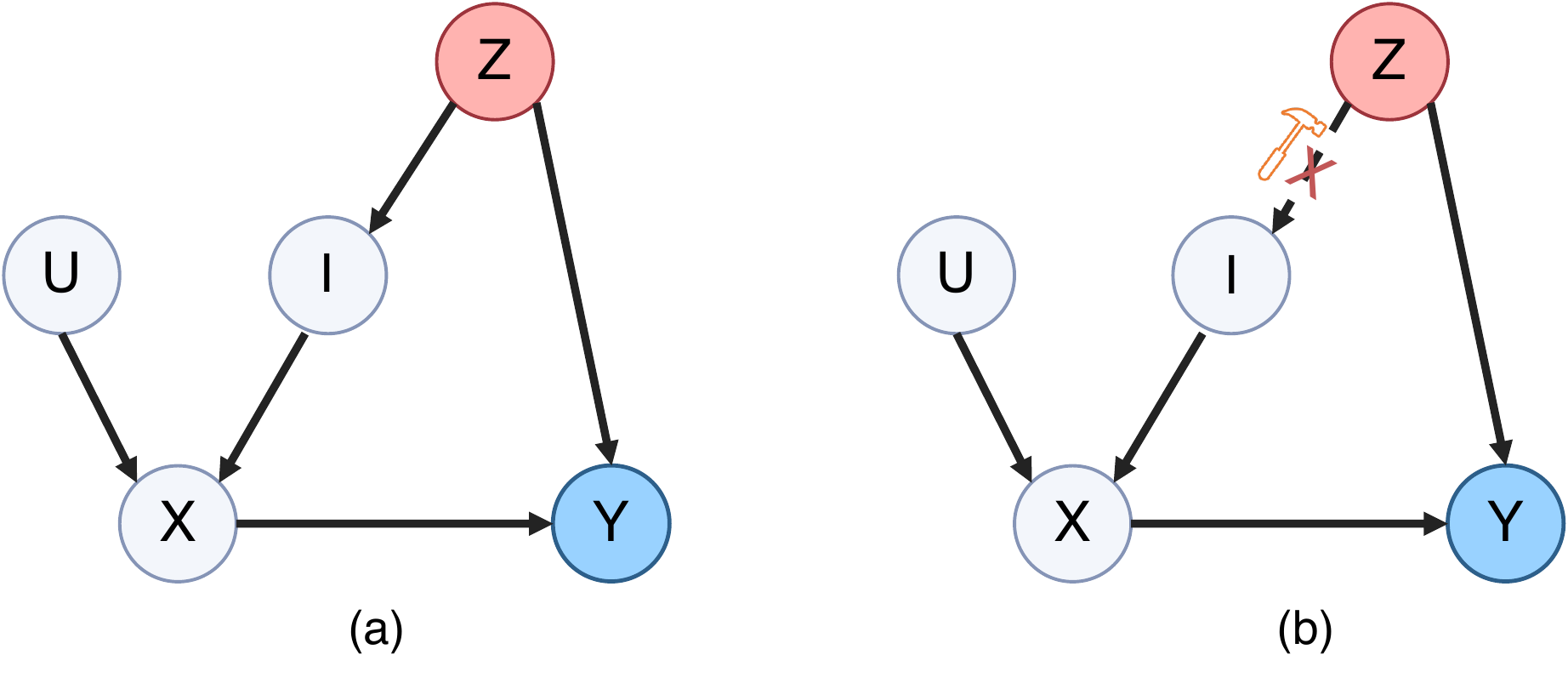}

\caption{
The illustration of causal interpretation in recommendations. 
$Z$ represents the hidden confounder and we primarily focus on underlying exposure, while $U$ and $I$ refer to the raw features of users and items. 
The representation of user-item pairs is represented by $X$, and $Y$ denotes the user's preference.
}
\label{506448163853}
\end{figure}

To further elaborate, We have abstracted a causal graph for the recommendation model in Figure \ref{506448163853}. 
In particular, the variables $Z, U, I, X$ and $Y$ represent specific elements of the model:
\begin{itemize}
    \item $Z$ denotes the confounder, which is unobserved or may be inaccessible for measurement. 

    \item $U$ represents raw features for the user, such as ID, gender and age. 
    \item $I$ refers to the item variable that includes raw features for the item, such as ID, colour and category.
    \item $X$ is the concatenation of user and item representations.

    \item $Y$ stands for the user preference variable.
\end{itemize}

Now we provide a comprehensive interpretation of subgraphs depicted in Figure \ref{506448163853}.  

$\{Z\rightarrow I\}$. The confounder may lead to the observation of only a subset of items, resulting in spurious correlations between the user and specific items. 
As demonstrated in Figure \ref{266492058979}, the user is interested in photography but due to the confounder, they can only see the computer. 
Through observational feedback, we infer that his preference may appear to be the computer, but their true interest may be obscured by the confounder.

$\{U,I \rightarrow X \rightarrow Y\}$. 
We note that the classic recommendation model follows this path. 
Raw features for observed users and items are fed into encoder networks, and the classifier outputs the binary ratings based on the concatenation of users' and items' embeddings.

$\{Z \rightarrow Y\}$. 
We also connect $Z$ to the outcome label $Y$. 
This ensures that the confounding impact of $Z$ on $Y$ is captured, other than the path between $U, I$ and $Y$.
For instance, the geographic location of users may impact the availability and appeal of the movies they can access, and subsequently affect their preferences \cite{wang2020causal}.
This also aligns with implicit exposure.  

\begin{remark}
\label{846056556858}
It is important to note that the confounder $Z$ can represent multiple possible influences on the observations $\{U,I \rightarrow X \rightarrow Y\}$ \cite{zhang2021causal, wang2021deconfounded, wei2021model}. 
However, this work primarily concentrates on the underlying exposure \cite{Gupta2021CorrectingEB} and presents a causality-aware interpretation.  

\end{remark}
The standard recommendation models estimate the conditional probability $P(Y\vert U, I)$ or $P(Y\vert X)$ based on observational feedback via the path $\{U,I \rightarrow X \rightarrow Y\}$. 
However, we can distinguish another path from $X$ to $Y$, which is $\{X \leftarrow Z \rightarrow Y\}$. 
According to the d-separation in \cite{pearl2009causal}, the path $\{X \leftarrow Z  \rightarrow Y\}$ is not blocked since it only contains chain and fork paths. 
Therefore, in order to obtain a more accurate inference for $Y$, it is necessary to block the path $\{Z \rightarrow I\}$.

In order to address the path $\{Z \rightarrow I\}$, we must implement solutions that effectively block the causal influence of the variable $Z$ on $I$.
However, it is often impractical to conduct actual interventions, such as randomized controlled trials, due to various constraints. 
Alternatively, causality provides do-calculus to calculate the intervention $P(Y\vert do(X))$ and a key method named backdoor criterion. 

It is necessary to have knowledge of the prior distribution of the confounder $p(Z)$ and solve the sum operator. 
Previous studies have proposed methods for analyzing the confounder by stratifying it into pieces \cite{zhang2020causal, deng2021comprehensive}, or by encoding the intervention as a new variable \cite{yang2019causal}. 
However, in the context of recommender systems, the confounder $Z$ refers to an unknown exposure mechanism, making it difficult to decompose and stratify. 

Therefore, an alternative approach is to  simulate the intervention in representation space \cite{mitrovic2020representation, yue2020interventional}.

The randomized controlled experiment allows us to infer that a user is unlikely to exhibit a preference for an item if there is no observable interaction. 
To further enhance our understanding of this phenomenon, we consider the use of contrastive SSL due to its advanced capabilities for representation learning, and present an analysis on how to leverage it for reducing exposure bias.

\subsection{Exposure Bias Reduction}
\label{848603119898}

We propose to address exposure bias, which is the problem of models being trained on a limited subset of the data, resulting in poor generalization to new data.
We address this within a contrastive self-supervised learning framework.
This framework leverages data augmentation operators to generate positive samples.
It also optimizes encoders by maximizing agreements between the anchor and the positive samples. 

This approach is in line with the positive sampling method $p_{pos}$ outlined  in \cite{wang2020understanding,chen2020simple}:
\begin{equation}
    \label{810933119091}
    \mathcal{L}_{con} \triangleq \frac{1}{\vert \mathcal{D} \vert} \sum_{\substack{(u,i)\in \mathcal{D}\\ i^{+} \sim p_{pos} \\ i^{-} \sim p_{data}}} -log \frac{exp(\phi_{\theta} (u,i^{+}))}{exp(\phi_{\theta} (u,i^{+})) + \sum_{i^{-}} exp(\phi_{\theta}(u, i^{-})  ) }
\end{equation}
where $\mathcal{D}$ represents the training dataset, $i^+$ represents a positive sample obtained using the positive sampling method $p_{pos}$ and $i^-$ refers to uniformly sampling from the data distribution.
$\phi_{\theta}(\mathbf{u}, \mathbf{v})$ measures the similarity between two vectors of $\mathbf{u}$ and $\mathbf{v}$. 
This is achieved through the use of cosine similarity and the incorporation of a scalar temperature hyper-parameter $\tau$.

Additionally, it should be noted that during each mini-batch, negative samples are selected from the remaining samples that do not include the anchor and positive ones. 
To clarify, negative samples are identified after the positive samples have been determined. 
Let $C = \{i^{+}\}$ represent the set of positive samples, and $C^{-} = \{i^{-}\} = C \cup \{i\}$ correspond to the set of negative samples.
We define $p_{\theta}^C$ to represent the probability distribution over the set of positive samples  $C$, given a user-item pair $(u,i)$ and a model parameterized by $\theta$.
We subsequently integrate the positive sampler into Eq.(\ref{810933119091}) by combining it with the set $C$:

\begin{equation}
\begin{aligned}
    \label{791901700105}
    \mathcal{L}_{con} &\propto \frac{1}{\vert \mathcal{D} \vert} \sum_{\substack{(u,i)\in \mathcal{D}\\ i^{+} \in C }} - p_{pos}(i^{+} \vert u,i)  log \frac{exp(\phi_{\theta} (u,i^{+}))}{\sum_{i^{-} \in C^{-}} exp(\phi_{\theta}(u,i^{-})) } \\
    &= \frac{1}{\vert \mathcal{D} \vert} \sum_{(u,i) \in \mathcal{D}} - p_{pos} log p_{\theta}^{C^{-}} \\
    & \overset{p_{pos}:=\frac{1}{P_{u,i}}}{=} \frac{1}{\vert \mathcal{D} \vert} \sum_{(u,i) \in \mathcal{D}} - \frac{1}{P_{u,i}} log p_{\theta}^{C^{-}} \\
    & \overset{C^{-}:= \mathcal{I}}{=} \frac{1}{\vert \mathcal{D} \vert} \sum_{(u,i) \in \mathcal{D}} - \frac{1}{P_{u,i}} log p_{\theta} 
\end{aligned}
\end{equation}

Then, we compare the loss function of Eq.(\ref{791901700105}) to that of inverse propensity scored-based methods:
\begin{equation}
\label{778293088365}
    {\mathcal{L}}_{IPS} =  \frac{1}{\vert {(u,i)\in \mathcal{D} \vert }}\sum_{(u,i)\in \mathcal{D}: O_{u,i}=1} - \frac{log p_{\theta}(i \vert u)}{P_{u,i}}
\end{equation}   
where 
\begin{equation}
\label{477420454949}
    p_{\theta}(i \vert u) = \frac{exp(\phi_{\theta} (u,i))}{\sum_{i^{'} \in \mathcal{I}} exp(\phi_{\theta}(u, i^{'}))}
\end{equation}  

\begin{remark}
\label{940921303746}
The Eq.(\ref{791901700105}) underscores the potential of contrastive self-supervised learning (SSL) to mitigate exposure bias, a key challenge impacting the interpretability and fairness of recommender systems.
This approach aligns with findings from inverse propensity score (IPS)-based studies. 
Notably, the derivation suggests that positive sampling in contrastive SSL can be achieved through the use of inverse propensity scores. 
Furthermore, it is recommended to expand and diversify the set of positive samples $C$ so that it closely mirrors $I$.

\end{remark}

While Remark \ref{940921303746} is similar to \cite{zhou2021contrastive} in terms of utilizing negative examples through item popularity probability $p_{data}(i)$ for deep candidate generation in recommender systems, our focus is on the positive samples in a popular contrastive SSL framework \cite{chen2020simple}. 
In addition, we propose and empirically evaluate a strategy for expanding and diversifying the positive sample set $C$ as a complement to this approach.

\subsection{Contrastive Counterfactual Learning}
\label{068927995050}

We present a analysis of  the asymptotic consistency of contrastive learning in relation to inverse propensity score methods, by performing positive sampling from the inverse propensity scores or expanding the set of positive samples to the entire item set. 
Building on this finding, we propose the corresponding positive sampling methods related to counterfactual inference for contrastive SSL, namely contrastive counterfactual learning (CCL) as depicted in Figure \ref{927512613774}(b) and (c). 

\subsubsection{Contrastive Counterfactual Learning (CCL)} 
We generate positive samples through the use of data augmentation operators and optimize encoders by maximizing the agreements between the anchor (a selected data example) and the positive samples. 
Specifically, given a mini-batch samples with $N$ examples, we use the concatenation $\mathbf{x}_k$ of the user and item's features $\mathbf{x}_u \oplus \mathbf{x}_i$ to obtain $\{\mathbf{x}_k, \hat{y}_k\}_{k=1}^N$. 
We then propose three informative samplers ($\mathcal{T}$) related to counterfactual inference and select one of them ($t\sim \mathcal{T}$) to transform the mini-batch samples. 

We obtain a total of $2N$ samples with $\tilde{\mathbf{x}}_{2k-1}=\mathbf{x}_k$ and $\tilde{\mathbf{x}}_{2k}=t(\mathbf{x}_k)$ as follows:
\begin{equation}
\{\mathbf{x}_1,\tilde{\mathbf{x}}_1, \mathbf{x}_2,\tilde{\mathbf{x}}_2,...,\mathbf{x}_k,\tilde{\mathbf{x}}_k,...,\mathbf{x}_N,\tilde{\mathbf{x}}_N\}    
\end{equation}
where $k\in \{1,2,...,N\}$. 
The anchor and the chosen positive sample $(\mathbf{x}_k,\tilde{\mathbf{x}}_k)$ are considered as positive pairs, and we treat the remaining $2(N-1)$ samples as negative samples. 
\begin{equation}
\label{l_ssl}
    l_{ccl}(\tilde{\mathbf{h}}_{2k-1},\tilde{\mathbf{h}}_{2k})=-log\frac{exp(sim(\tilde{\mathbf{h}}_{2k-1},\tilde{\mathbf{h}}_{2k})/\tau)}{\sum_{m=1}^{2N}\mathbbm{1}_{[m\neq 2k-1]}exp(sim(\tilde{\mathbf{h}}_{2k-1},\tilde{\mathbf{h}}_{m})/\tau))}
\end{equation}
where $\mathbbm{1}_{[m\neq 2k-1]} \in \{0,1\}$ is an indicator function and the indicator equals to $1$ iff $m\neq 2k-1$. 
Let $sim(\mathbf{u}, \mathbf{v})$ represent the dot product for measuring similarity between the vector $\mathbf{u}$ and $\mathbf{v}$. And $\tau$ denotes the temperature parameter. The complete CCL loss is defined as follows:  
\begin{equation}
\label{loss_total_ccl}
    \mathcal{L}_{ccl} = \frac{1}{2N} \sum_{k=1}^{N}[ l_{ccl}(\tilde{\mathbf{h}}_{2k-1},\tilde{\mathbf{h}}_{2k})+l_{ccl}(\tilde{\mathbf{h}}_{2k},\tilde{\mathbf{h}}_{2k-1})]
\end{equation}

\subsubsection{Three Novel Positive Samplings}
\label{sec_pos}

We propose the implementation of three novel positive samplers for classic contrastive SSL in order to mitigate confounding bias introduced by the underlying exposure system, as illustrated in Figure \ref{927512613774}(c). 

\begin{itemize}[leftmargin=*]
    \item \textbf{Propensity score-based sampling}. 
    The propensity score-based sampling is motivated by \cite{zhou2021contrastive} which sets the negative sampler as the propensity score function for debiased candidate generation. 
    As Eq.(\ref{653054168312}) shows, the propensity score $P_{u,i}$ estimates exposure probability for the user-item pair. 
    To manipulate uniform exposure, we select the user-item pair with the most significant difference in propensity value from the current sample, while keeping the user constant. 
    For example, as illustrated in Figure \ref{927512613774}(c), we randomly select a user-item pair, such as $(u_2, i_2)$, as the anchor from the matrix. 
    Next, we calculate the propensity score vector for the user $u_2$.
    We sort the elements in the vector and select the item with the greatest difference in propensity score from the current user-item pair, which in the case is $(u_2, i_4)$. 
    There are two methods for estimating the propensity score \cite{schnabel2016recommendations}: one is through a naive Bayes estimator with a small set of unbiased data, and the other is logistic regression using the raw features of users and items.    
    
    \item \textbf{Item popularity-based sampling}. 
    This method addresses the challenge of estimating the propensity score precisely by using item popularity instead. 
    To ease estimation and avoid small propensities, some works use the item popularity in replace of propensity scores  \cite{zhou2021contrastive, liu2021contrastive}. 
    To this end, we propose the use of item popularity as a sampling method in contrastive SSL. 
    Specifically, we determine the item popularity by counting the number of interactions per item, which is equivalent to summing each column in the indicator matrix $O$. 
    Next, we find the maximum number of interactions for each item and divide the count by the maximum. 
    Finally, we use the root square of the resulting value as the item popularity. 
    Once estimating the item popularity, we sort the items by popularity and select the one with the most significant difference in popularity from the current item. 
    This results in the selection of positive samples through item popularity. 
    For example, using this method, we would choose $(u_2, i_1)$ as the positive sample for the anchor $(u_2, i_2)$ in Figure \ref{927512613774}(c).  
    
    \item \textbf{Random counterfactual sampling}. 
    Unlike traditional random operators such as cropping or masking used in sequential recommender systems \cite{xie2020contrastive, zhou2020s3}, our random sampling method focuses on matching unexposed items to users. 
    For example, in Figure \ref{927512613774}(c), we can see a toy user-item matrix with five users and items, where gray squares represent interacted user-item pairs without interactions and other squares represent interactions. 
    We treat $(u_2, i_2)$ as the anchor, and $(u_2, i_0)$ cab be considered as a positive sample since it is an uninteracted (counterfactual) pair for the user $u_2$.
    The goal of this technique is to make items as evenly exposed to a user in the representation space, thus excavating users' true preferences. 
    Interestingly, this simple sampling method from random counterfactual samples leads to superior performance.
    
\end{itemize}
In conclusion, the three sampling methods presented are both practical and effective. 
They employ either an estimated exposure probability or widely available counterfactual samples, aligning with the previous finding of theoretical remark \ref{940921303746}. 
We will evaluate and analyze the performance of these three methods in Section \ref{774332982972}.

\subsection{Training and Optimization}
As illustrated in Figure \ref{927512613774}(a), the overall training process comprises of two main components: a contrastive learning module with $\mathcal{L}_{ccl}$ and a multi-layers prediction module with $\mathcal{L}_{rec}$. 
We use a multi-task training approach to optimize these two loss functions as follows:
\begin{equation}
\label{total_loss}
    \mathcal{L}=\mathcal{L}_{rec}+\lambda \mathcal{L}_{ccl}
\end{equation}
where $\lambda$ is a hyper-parameter for controlling the weight of contrastive counterfactual learning. 
We summarize the training process in Algorithm \ref{alg1}. 
Alternatively, it is also possible to pre-train the $\mathcal{L}_{ccl}$ and then fine-tune $\mathcal{L}_{rec}$.

\begin{algorithm}

\renewcommand{\algorithmicrequire}{\textbf{Input:}}
\renewcommand{\algorithmicensure}{\textbf{Output:}}
\caption{\bf{:} Training process for CCL}
\label{alg1}
\begin{algorithmic}[1]
	\State \textbf{Input:} Observed feedback data $\mathcal{D}$, hyper-parameters: weight of the CCL module $\lambda$, temperature $\tau$, batch size $N$, the embedding size for $g_u$ and $g_i$, the number of hidden layers in $f$.
	\State \textbf{Output:} $f, g_{u}, g_{i}$ 
	\Repeat
	    \For{mini-batches $\{(\mathbf{x}_k, \hat{y}_{k})\}_{k=1}^{N}$ sampled from $\mathcal{D}$}
	        \LineComment{$\tilde{\mathbf{x}}_{k}$ is the concatenation of the embeddings $\mathbf{x}_u$ and $\mathbf{x}_i$}
	        \For {all $k \in \{1,2,...,N\}$ }
	        \State{draw the sampling method $t$ from three proposed samplings}
	        \State{$\tilde{\mathbf{x}}_{2k-1}=\mathbf{x}_k$}
	        \State{$\mathbf{h}_{2k-1}=g_u(\mathbf{x}_u)\oplus g_i(\mathbf{x}_i)$}
	        \State{$\tilde{\mathbf{x}}_{2k}=t(\mathbf{x}_k)$}
	        \State{$\mathbf{h}_{2k}=g_u(\tilde{\mathbf{x}}_u)\oplus g_i(\tilde{\mathbf{x}}_i)$}
	        
	        \EndFor
	        \State{\textbf{end for}}
	       \State{Calculate the contrastive counterfactual loss $\mathcal{L}_{ccl}$ with Eq.(\ref{loss_total_ccl}) }
	        
	        \State{$\mathcal{L}_{rec}=\sum_{k=1}^{N}l_{rec}(\mathbf{h}_{2k-1})$ with Eq. (\ref{l_rec})}
	        
	        \State{Calculate the total loss $\mathcal{L}$ with Eq. (\ref{total_loss})}
	    \State{Update the networks $f, g_u, g_i$ to minimize $\mathcal{L}$}
	    \EndFor
	    \State {\textbf{end for}}
	\Until{convergence}
	
\end{algorithmic}  
\end{algorithm}

\begin{table*}[t]
\caption{Performance Evaluation on top-n ranking metrics. We report results for NDCG, Recall, MRR, Gini and Global utility on Coat and Yahoo! R3, with the best performance denoted in bold and the second best with an underline.
}
\label{907914275759}
\begin{tabular}{c l c c c c c c c}
\hline
Dataset                & Methods    & NDCG@5 $\uparrow$   & NDCG@10 $\uparrow$  & Recall@1 $\uparrow$ & Recall@5 $\uparrow$ & MRR $\uparrow$  & Gini $\downarrow$ & Global Utility $\uparrow$   \\ \cline{1-9} 
\multirow{11}{*}{Coat} & MF   & 0.618878 & 0.685805 & 0.143029 & 0.470560 & 0.724109 & {0.328557} & \underline{0.519310} \\
                       & +IPS    & 0.546103 & 0.642255 & 0.137241 & 0.445300 & 0.695958 & 0.370406 & 0.465517 \\
                       & +SNIPS  & 0.619469 & 0.693272 & 0.130670 & 0.472532 & 0.702422 & 0.335425 & 0.509655  \\
                       & +DR     & 0.610190 & 0.683095 & 0.142121 & 0.458575 & 0.723581 & 0.335808 & 0.506207 \\
                       & +CVIB   & 0.635618 & 0.706618 & 0.154944 & 0.485840 & \underline{0.753408} & 0.333855 & 0.515172\\ \cline{2-9}
                       & NCF  & 0.613597 & 0.686848 & 0.155841 & 0.478001 & 0.727774 & 0.356257 & 0.502483 \\
                       & +IPS   & 0.615375 & 0.692377 & 0.149823 & 0.469947 & 0.726316 & 0.347815 & 0.506897 \\
                       & +SNIPS & 0.619516 & 0.697859 & {0.156743} & 0.464328 & 0.739233 & 0.343668 & 0.509655 \\
                       & +DR    & \underline{0.636087} & \underline{0.708855} & 0.152474 & 0.469774 & 0.726123 & 0.331799 & 0.519310 \\
                       & +CVIB  & 0.627515 & 0.702538 & 0.155350 & \underline{0.488107} & 0.742603 & 0.341349 & 0.5155862 \\ \cline{2-9} 
                       & LightGCN & 0.598313 & 0.671740 & 0.133139 & 0.454676 & 0.698339 & 0.358796 & 0.502759 \\ \cline{2-9} 
                       & CCL w/ps & 0.631428 & 0.705370 & \textbf{0.158229} & 0.478170 & 0.748094 & \underline{0.325380} & 0.513103 \\
                       & CCL w/pop & 0.614438 & 0.690041 & 0.136208 & 0.461086 & 0.735151 & 0.330061 & 0.513103 \\
                       & CCL w/cf       & \textbf{0.646099} & \textbf{0.715099} & \underline{0.156814} & \textbf{0.490682} & \textbf{0.755027} & \textbf{0.312651} & \textbf{0.524828} \\ \hhline{ = = = = = = = = = }

\multirow{11}{*}{Yahoo! R3} & MF   & 0.634687 & 0.762871 & 0.381295 & 0.667766 & 0.436356 & 0.582162 & 0.252037 \\
                        & +IPS    & 0.646120 & 0.765412 & 0.373860 & 0.700561 & 0.430910 & 0.557144 & 0.259889 \\
                        & +SNIPS  & 0.638394 & 0.763969 & 0.381063 & 0.683709 & 0.432395 & 0.565564 & 0.251926 \\
                        & +DR     & 0.656221 & 0.772937 & 0.385328 & 0.704601 & 0.444278 & 0.557553 & 0.263815 \\
                        & +CVIB   & \underline{0.696131} & \underline{0.799171} & \underline{0.412612} & \underline{0.738778} & \underline{0.483970} & \underline{0.541144} & \textbf{0.278963} \\ \cline{2-9}
                        & NCF  & 0.654634 & 0.774168 & 0.383378 & 0.700092 & 0.443718 & 0.562444 & 0.260481 \\
                        & +IPS   & 0.660373 & 0.775258 & 0.381428 & 0.710659 & 0.436361 & 0.548205 & 0.263970 \\
                        & +SNIPS & 0.652109 & 0.768611 & 0.390112 & 0.724687 & 0.452089 & 0.549243 & 0.261852 \\
                        & +DR    & 0.647838 & 0.768016 & 0.387093 & 0.718626 & 0.452037 & 0.559978 & 0.259259 \\
                        & +CVIB  & 0.667489 & 0.780986 & 0.394774 & 0.716073 & 0.456541 & 0.547794 & 0.2655923 \\ \cline{2-9}
                        & LightGCN & 0.589551 & 0.730583 & 0.343585 & 0.645705 & 0.376967 & 0.582675 & 0.235222 \\ \cline{2-9} 
                        & CCL w/ps & 0.685223 & 0.792537 & 0.405060 & 0.730238 & 0.475835 & 0.544201 & 0.273185 \\
                        & CCL w/pop & 0.685183 & 0.792550 & 0.405038 & 0.730168 & 0.475966 & 0.544143 & 0.273111 \\
                        & CCL w/cf       & \textbf{0.696671} & \textbf{0.799688} & \textbf{0.413482} & \textbf{0.739955} & \textbf{0.487297} & \textbf{0.537753} & \underline{0.277519} \\  \cline{1-9}

\end{tabular}
\end{table*}

\section{Experiments}
\label{sec_experiment}

In this section, we evaluate our model using two real-world datasets, Coat and Yahoo! R3. 
These datasets have been specified as benchmarks in prior research. 
We not only evaluate the performance but also conduct an in-depth examination of different positive sampling strategies, provide visualization studies, and present an ablation study.  

\subsection{Dataset Description}
\textbf{Coat\footnote{https://www.cs.cornell.edu/~schnabts/mnar/}.} \cite{schnabel2016recommendations} first introduces the Coat dataset on recommendations, which consists of customer data for shopping for coats on an online website. 
The total user-item matrix includes 290 users and 300 items. 
The training data includes each user rating 24 coats with self-selection and 16 randomly displayed coats for the test set. 
The ratings are on a five-star scale, and we have binarized them so that rates greater than or equal to three are regarded as positive feedback and those smaller than three are considered negative. 

\noindent \textbf{Yahoo! R3\footnote{https://webscope.sandbox.yahoo.com/catalog.php?datatype=r}.}
This dataset is correlated with user-song ratings \cite{marlin2009collaborative}. 
The training set includes more than 300K ratings from 15400 users who made selections. 
The testing set contains ratings from a subset of 5400 users who were asked to rate ten songs at random.

The Coat and Yahoo datasets, which consist of both self-selected training data and unbiased test data with random exposure, exhibit varying data sizes. 
Through the utilization of these datasets in experiments, it is proposed that robust support for model scalability is present in mitigating confounding biases and inferring accurate user preferences is demonstrated.

\subsection{Experimental Setting}
\label{sec_experimental_setting}

\textbf{Baselines.} We compare our CCL method with several state-of-the-art baselines. 1) Base models: matrix factorization (MF) \cite{koren2008factorization}, neural collaborative ﬁltering (NCF) \cite{he2017neural} and LightGCN \cite{He2020LightGCNSA}. 
2) Propensity score-based methods: Inverse Propensity Scoring (IPS) \cite{schnabel2016recommendations}, Self Normalized Inverse Propensity Scoring (SNIPS) \cite{swaminathan2015self}, and the Doubly robust method that combines imputed errors and propensities \cite{wang2019doubly}. 
3) Side information-based techniques that require unbiased data or network information. 
But as the two datasets are not correlated with the social network, we follow the procedure in \cite{schnabel2016recommendations} and adopt a small sample of MCAR data for propensity estimation in the Coat dataset. 
Thus, we regard IPS on Coat dataset as the side-information-based method. 
4) Information bottleneck-based methods: counterfactual variational information bottleneck (CVIB) \cite{wang2020information} which utilizes information-theoretic representation learning to learn a balanced model between the factual and counterfactual domains. 

It is crucial to acknowledge that the studies conducted by \cite{Zhang2021CausalIF,Wei2020ModelAgnosticCR} can not be directly compared to our approach, as they focus on a different aspect of bias, specifically popularity bias, which is a inherent property rather than stemming from the underlying exposure mechanism.
Additionally, their experiments do not involve evaluating the model on a dataset that is randomly interacted with by users.
It is also noteworthy that the various methods within the aforementioned categories can be integrated as plugins into the two base models.

\textbf{Metrics.} We use two types of evaluation metrics as recommended in \cite{schnabel2016recommendations, bonner2018causal}: rating prediction metrics, such as Mean Average Error and AUC, and top-n raking metrics, including NDCG, Recall, MRR, Gini and Global utility \cite{carraro2022sampling}. 

\textbf{Implementation Details.} 
We utilize the code provided by the  CVIB\footnote{https://github.com/RyanWangZf/CVIB-Rec} model and set all hyper-parameters as reported in the baseline papers. 
We implement our model using PyTorch and set the hyper-parameters within the ranges of \{4, 8, 16, 32, 64\} for embedding size, \{1, 2, 3\} for the number of hidden layers, and \{512, 1024, 2048, 4096\} for the batch size. 
Additionally, we select the weight of SSL $\lambda$, the temperature $\tau$, learning rate and weight decay within the ranges of [0.5, 1.5], [0.1, 1.5], \{$1e$-5, $1e$-4, $1e$-3, $1e$-2\} and \{$1e$-4, $1e$-3, $1e$-2\} respectively. 
The model is optimized using the Adam optimizer. 
To address the long tail effect present in the training data, especially for Yahoo dataset, we also explore the use of binary-cross-entropy loss and focal loss for $\delta(\cdot)$ in Eq.(\ref{l_rec}).

\subsection{Performance Evaluation}
We evaluate our model and ten baselines on two benchmark datasets. 
We compute the mean of all metrics with 10 runs following the pioneer work \cite{wang2020information} and set the significance level at $0.05$ for all statistical tests on our method, as reported in Table \ref{907914275759} and \ref{364253059135}.

We first present results on seven top-n ranking metrics in Table \ref{907914275759}.
The evaluation on unbiased datasets allows us to estimate and prove the effectiveness of inferring natural users' preferneces.
The results in Table \ref{907914275759} show that our CCL model (includes three sampling methods) outperforms all baselines in terms of NDCG, Recall, MRR and Gini index on both benchmark datasets. 
Additionally, our method achieves the best global utility on Coat and the second-best result on the Yahoo dataset. 
These outstanding results further confirm the superiority of our method in generating unbiased recommendations. 
In particular, on the Coat dataset, our model surpasses a popular baseline, IPS, and the state-of-the-art method, CVIB, by $2.3\%$ and $1.3\%$ in NDCG@10, respectively. 
The overall outstanding performance of our method on the seven ranking metrics on two different datasets further affirms its effectiveness. 
It is worth noting that the considerable performance of CVIB and our method, especially on the Yahoo dataset, which is much larger than Coat, suggests that representation learning-based methods are more suitable for larger datasets.
Propensity scores may be competitive, but representation learning is a better solution for larger datasets.

We present the results of our method on two rating prediction metrics (MAE and AUC). 
As shown in Table \ref{364253059135}, our proposed method consistently outperforms all baselines on the Coat dataset and achieves the second-best results on Yahoo! R3 dataset. 
The IPS and SNIPS methods estimate propensity scores using naive Bayes with a small set of unbiased data on Yahoo and via logistic regression on the Coat dataset. 
However, despite not utilizing RCTs data, our approach still outperforms or matches these baselines, with a $3\%$ improvement in AUC on Yahoo! R3. 
Additionally, our method and CVIB show significant improvements in terms of AUC on Yahoo! R3 dataset. 
Compared to other baseline methods, MF-CVIB and ours achieve the top-2 best results and obtain 2\%-3\% improvements in terms of AUC. 
The AUC metric is a more robust measure of prediction classification \cite{Hanley1982TheMA}. 
Our results suggest that representation learning, specifically informational theoretic learning or contrastive SSL, is a good solution for unbiased recommendations, particularly for large datasets like Yahoo! R3.
This conclusion is consistent with the results on ranking metrics.

It is important to note that rating prediction and top-n ranking belong to different types of metrics and practical recommender systems often have to make a trade-off between them.
Typically, these systems focus on ranking metrics \cite{wang2020information}. 
We report the results with two different parameters in Table \ref{907914275759} and \ref{364253059135}.

\begin{table}[!ht]
\caption{Performance Evaluation of rating prediction. We report MAE and AUC for Coat and Yahoo! R3.
The results are highlighted in bold for the best and underlined for the second best.
}
\label{364253059135}

\begin{tabular}{l|c c|c c}
\hline
\multirow{2}{*}{Methods} & \multicolumn{2}{c}{Coat}  & \multicolumn{2}{c}{Yahoo! R3} \\ \cline{2-5}
                         & MAE $\downarrow$     & AUC $\uparrow$ & MAE $\downarrow$     & AUC $\uparrow$ \\  \cline{1-5}
MF                 & 0.058 &   0.702818     & 0.062 & 0.680819       \\
+IPS                  & 0.104 & 0.632157  & 0.050 & 0.683447       \\
+SNIPS                & 0.053 & 0.708457     & 0.037 & 0.682086       \\
+DR                   & 0.075 & 0.686495      & 0.046 & 0.687128       \\
+CVIB                 & 0.336 & 0.751686      & 0.515 & \textbf{0.710261}       \\ \cline{1-5}
NCF                & 0.046 & 0.7487968       & 0.111 & 0.677314       \\
+IPS                 & 0.053 & 0.7505703       & \underline{0.035} & 0.683527       \\
+SNIPS               & \underline{0.040} & 0.7533112       & \textbf{0.031} & 0.673665       \\
+DR                  & 0.041 & \underline{0.7682484}       & 0.054 & 0.677911       \\
+CVIB                & 0.218 & 0.7645173      & 0.117 & 0.692956      \\ \cline{1-5}
CCL w/ps & 0.097 & 0.726108 & 0.063 & 0.619925 \\
CCL w/pop & 0.088 & 0.736016 & 0.063 & 0.620073 \\
CCL w/cf & \textbf{0.040} & \textbf{0.773700}       & 0.072 &  \underline{0.702237} \\ \cline{1-5}

\end{tabular}
\end{table}

\subsection{In-depth Discussion}

\subsubsection{Ablation Study}

\begin{table}[b]
\caption{Ablation Study. $\mathcal{L}_{ccl}+\mathcal{L}_{rec}$ is our proposed model and the model with only $\mathcal{L}_{rec}$ represents the absence of the CCL module in Figure \ref{927512613774}.}
\label{table_ablation_study}
\begin{tabular}{c l | c c}
\hline
Dataset & Metrics & $\mathcal{L}_{ccl}+\mathcal{L}_{rec}$ & $\mathcal{L}_{rec}$ \\ \cline{1-4}
\multirow{9}{*}{Coat}  & MAE $\downarrow$ & \textbf{0.040} & 0.047 \\
                       & AUC $\uparrow$ & \textbf{0.774} & 0.761 \\
                       & NDCG@5 $\uparrow$   & \textbf{0.646} & 0.609\\
                       & NDCG@10 $\uparrow$   & \textbf{0.715} &0.690 \\
                       & Recall@1 $\uparrow$ & \textbf{0.157} & 0.153\\
                       & Recall@5 $\uparrow$ & \textbf{0.491} & 0.448 \\
                       & MRR $\uparrow$ & \textbf{0.755} & 0.738\\
                       & Gini $\downarrow$& \textbf{0.313} & 0.334\\
                       & Global Utility $\uparrow$ & \textbf{0.525} & 0.496 \\ \hhline{----}

\multirow{7}{*}{Yahoo! R3} 
                       & MAE $\downarrow$ & \textbf{0.072} & 0.076 \\
                       & AUC $\uparrow$ & \textbf{0.702} &  0.701\\
                       & NDCG@5 $\uparrow$   &  \textbf{0.697} & 0.596\\
                       & NDCG@10 $\uparrow$  & \textbf{0.800}  & 0.736 \\
                       & Recall@1 $\uparrow$& \textbf{0.413}  & 0.350\\
                       & Recall@5 $\uparrow$ & \textbf{0.740} & 0.647\\
                       & MRR $\uparrow$  & \textbf{0.487} & 0.387\\
                       & Gini $\downarrow$ & \textbf{0.538} & 0.585\\
                       & Global Utility $\uparrow$ & \textbf{0.278} & 0.237\\ \cline{1-4}
\end{tabular}

\end{table}

We conduct experiments to evaluate the contribution of the contrastive counterfactual module (CCL) in our model. 
We test the total loss $\mathcal{L}_{rec}+\mathcal{L}_{ccl}$, as well as only $\mathcal{L}_{rec}$ separately. 
The results, shown in Table \ref{table_ablation_study}, reveal that when combining the contrastive counterfactual module (with $\mathcal{L}_{ccl}$), we observe improved performance in all metrics, including rating prediction or top-n ranking metrics. 
It should be noted that, in this ablation study, we only present the best sampling method, which is random counterfactual sampling, for $\mathcal{L}_{ccl}$.

\subsubsection{Comparison of three positive samplings}
\label{774332982972}
\begin{figure}[!htbp]
\centering
\begin{subfigure}[t]{.48\linewidth}
  \centering 
  \includegraphics[width=\textwidth]{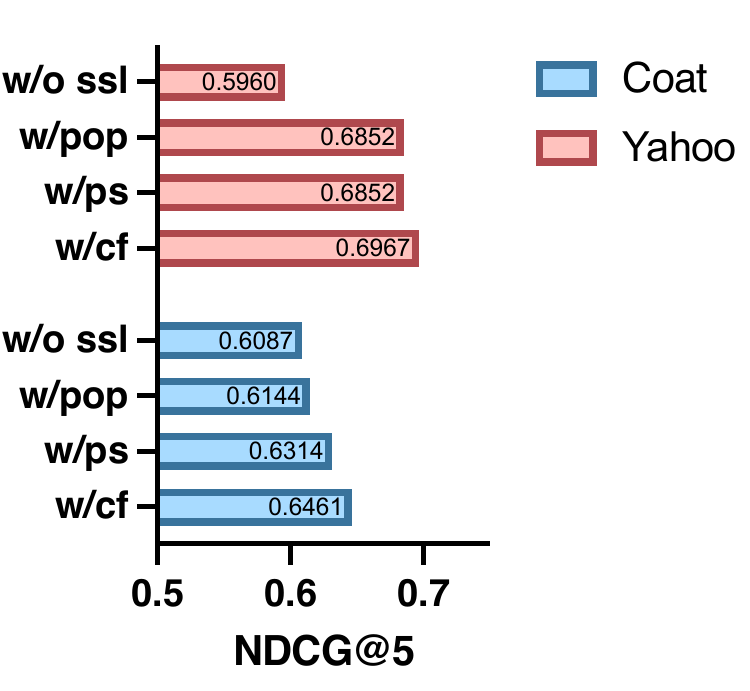}  
  \caption{NDCG@5}
  \label{fig:comprasion_ndcg_5}
\end{subfigure}
\hfill
\begin{subfigure}[t]{.48\linewidth}
  \centering

  \includegraphics[width=\textwidth]{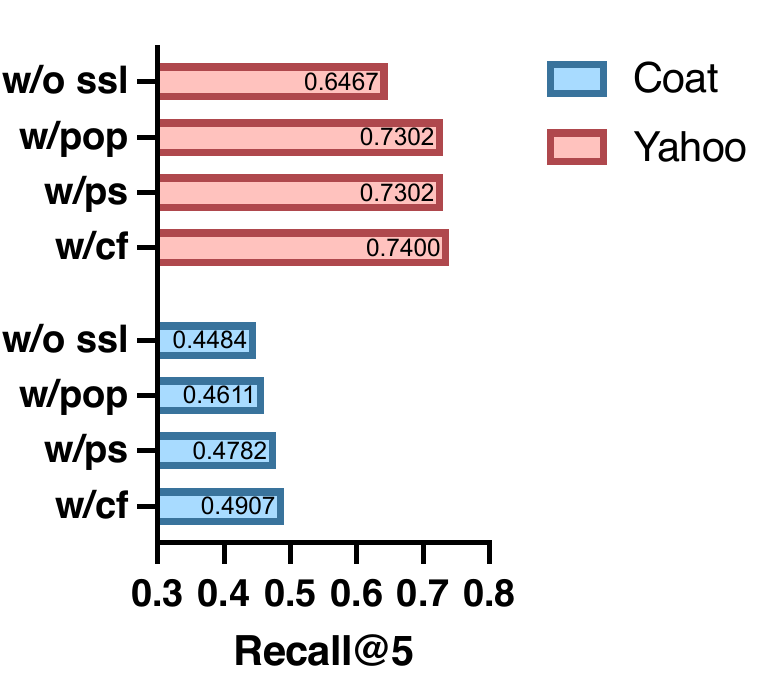}  
  \caption{Recall@5}
  \label{fig:comprasion_recall_5}
\end{subfigure}
\hfill

\caption{A comparison of performance, as measured by NDCG@5 and Recall@5, was conducted with respect to various sampling methods on both Coat and Yahoo datasets. 
The notation $'$w/$'$ and $'$w/o$'$ denote whether or not applying the sampling.}
\label{fig_comparison}
\end{figure}

The purpose of this experiment is to evaluate the effectiveness of three sampling strategies proposed in Section \ref{068927995050} and compare their performance. 
By conducting a comparative study on two benchmark datasets, we aim to explore the impact of these sampling methods and determine the optimal alternative. 
To accomplish this, we first construct a sampling set $\{$w/cf, w/ps, w/pop, w/o ssl$\}$ that includes random counterfactual, propensity score-based, item popularity-based and no contrastive SSL strategies. 
Figure \ref{fig_comparison} presents the comparison results of NDCG@5 and Recall@5 on both datasets. 
We observe that the performance consistently decreases on both datasets in the order of $\{$w/cf, w/ps, w/pop, w/o ssl$\}$. 
This proves that all three sampling methods contribute to significant performance improvements.
Additionally, we note that the performance with propensity score-based sampling is comparable to the item popularity-based method. 
However, estimating precise propensity scores can be difficult in practice, which may explain why some works choose to utilize item popularity as a substitute for propensity scores \cite{zhou2021contrastive}. 
What stands out in the figure is that the utilization of random counterfactual samples outperforms both the widely used propensity scores or item popularity. 
The empirical results presented in this study provide strong support for the assertion made in Remark \ref{940921303746} that expanding and diversifying the set $C$ of positive samples is beneficial.
Furthermore, as illustrated in Table \ref{907914275759} and \ref{364253059135}, both CVIB and our proposed method exhibit competitive performance in comparison to other baselines, particularly when applied to large-scale datasets, such as the Yahoo dataset.
It is noteworthy that both CVIB and our method utilize counterfactual samples,
which highlights the potential benefits of utilizing a vast number of counterfactual samples that are commonly overlooked in many studies. 

\begin{figure*}[t]
    \centering
    \begin{subfigure}[t]{.24\linewidth}
      \centering
      \includegraphics[width=\textwidth]{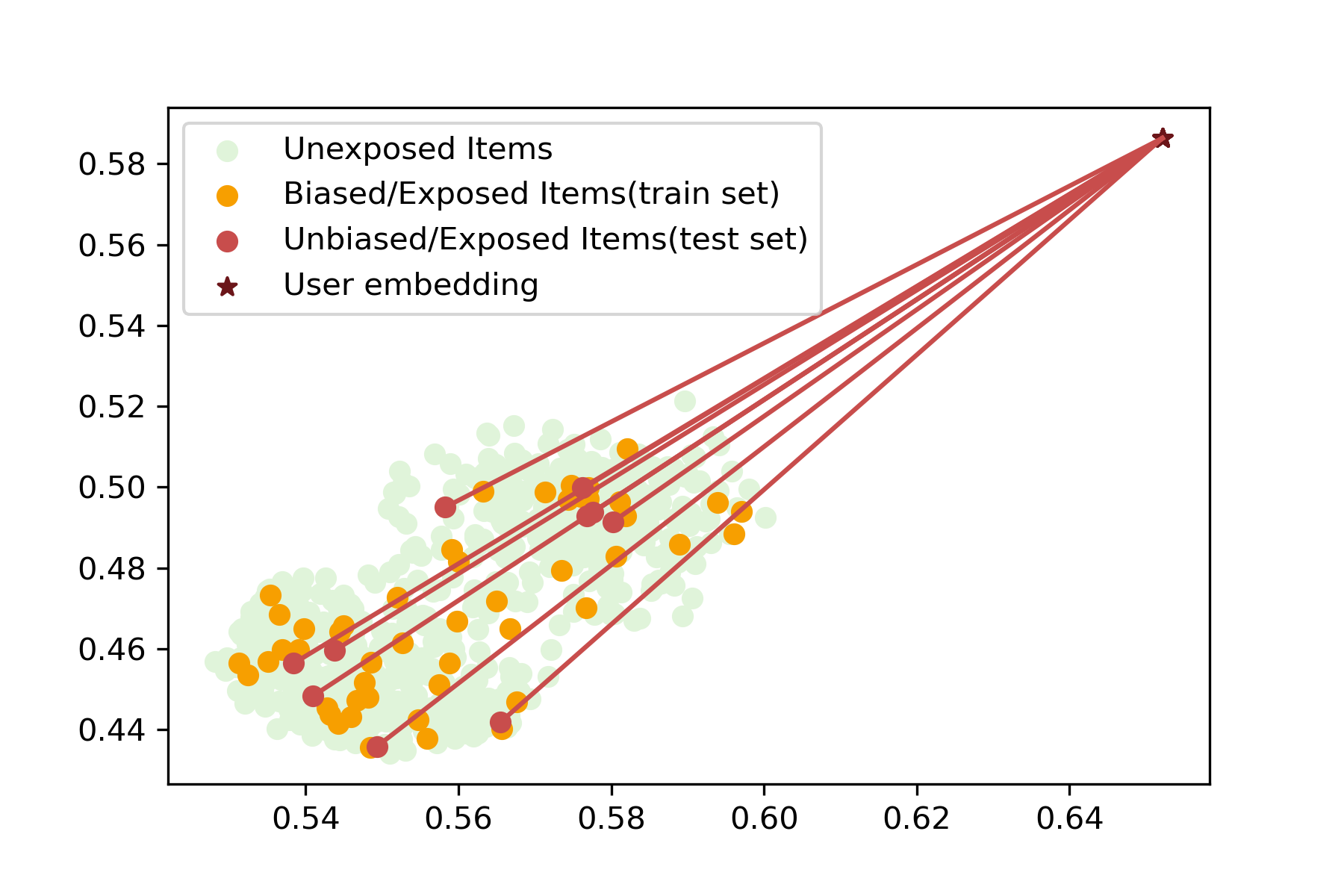}  
      \caption{CCL w/pop}
      \label{359247339191}
    \end{subfigure}
    \hfill
        \begin{subfigure}[t]{.24\linewidth}
      \centering
      \includegraphics[width=\textwidth]{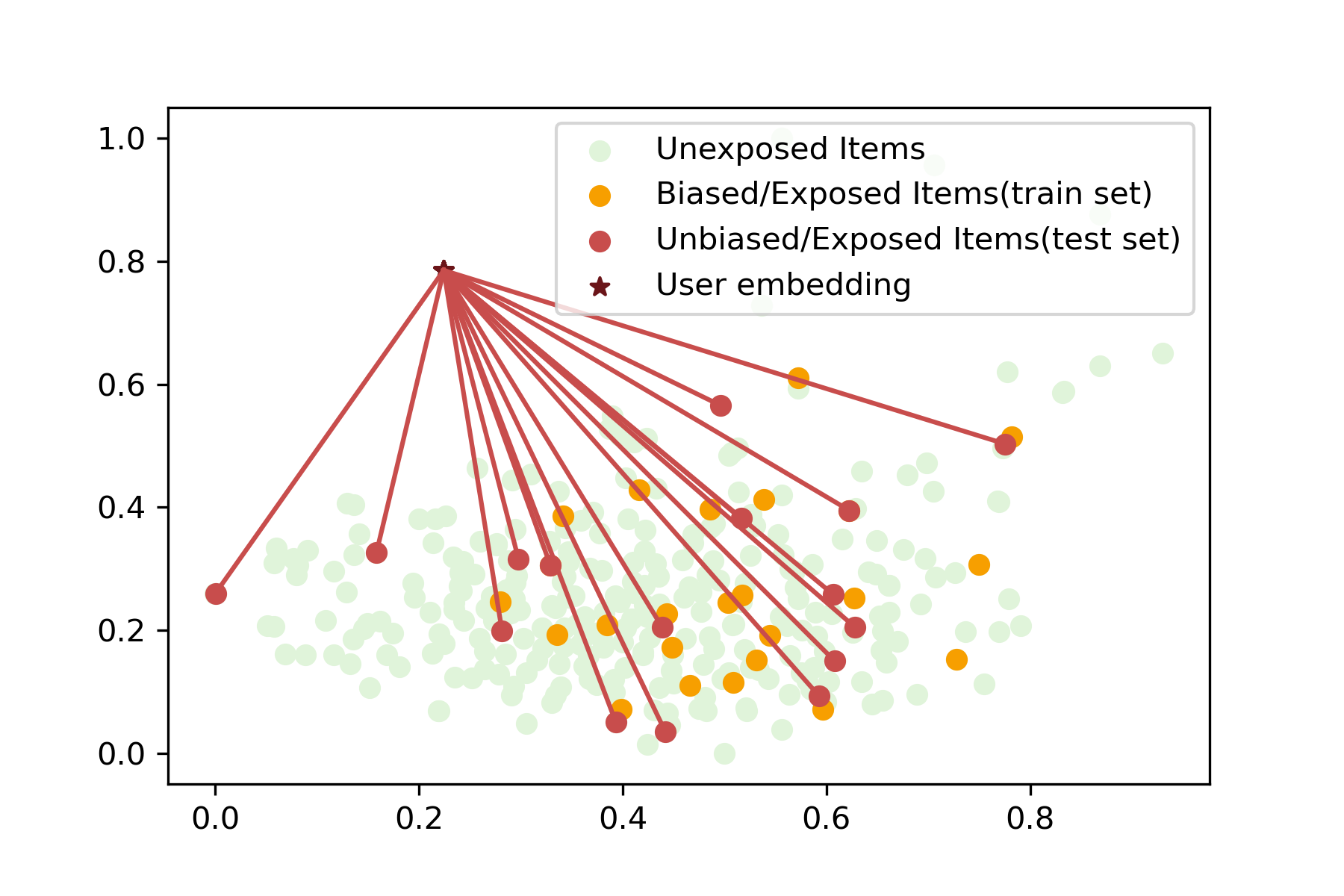}  
      \caption{CCL w/ps}
      \label{940490944135}
    \end{subfigure}
    \hfill
    \begin{subfigure}[t]{.24\linewidth}
      \centering
      \includegraphics[width=\textwidth]{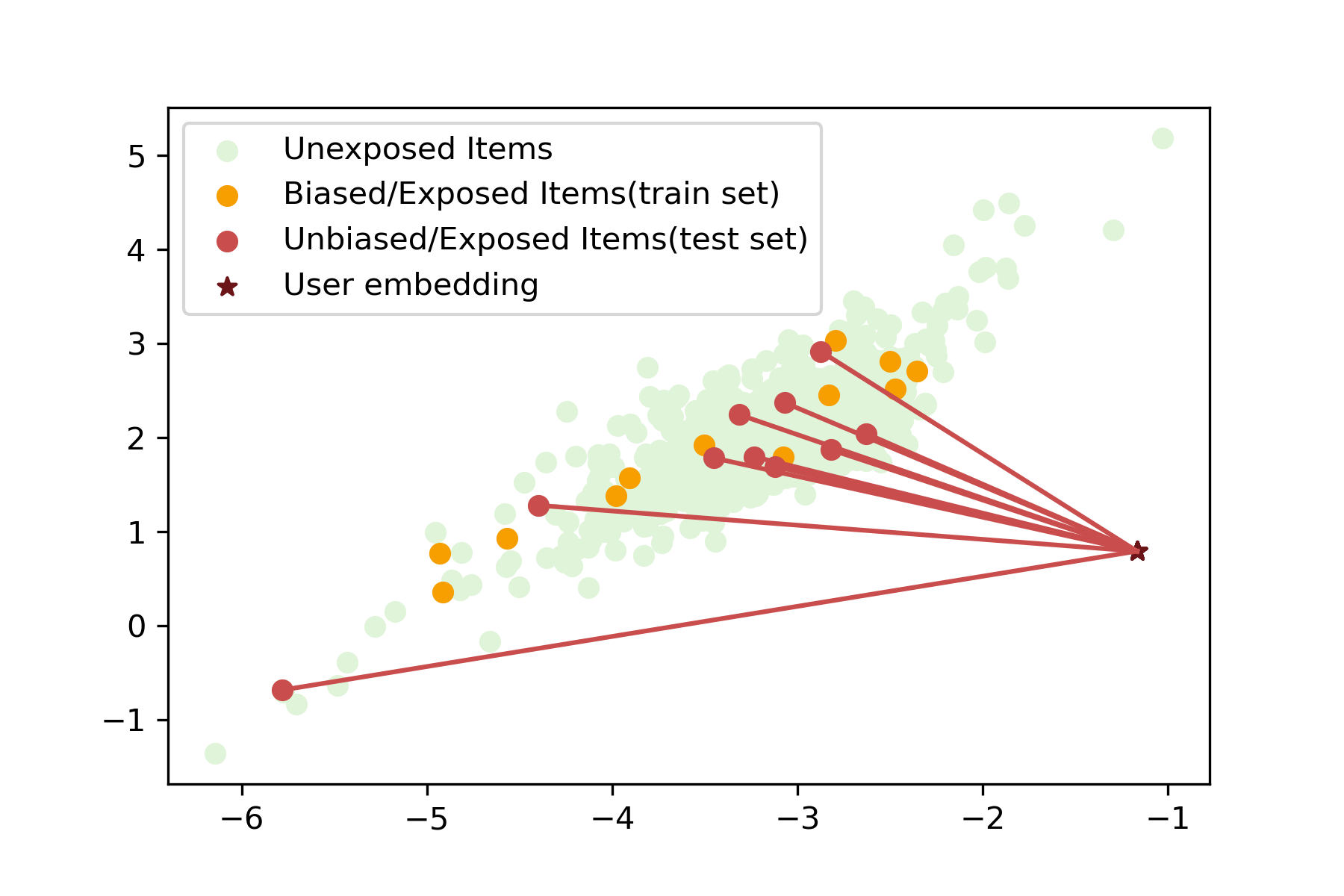}  
      \caption{MF\_CVIB}
      \label{fig:e}
    \end{subfigure}
    \hfill
    \begin{subfigure}[t]{.24\linewidth}
      \centering
      \includegraphics[width=\textwidth]{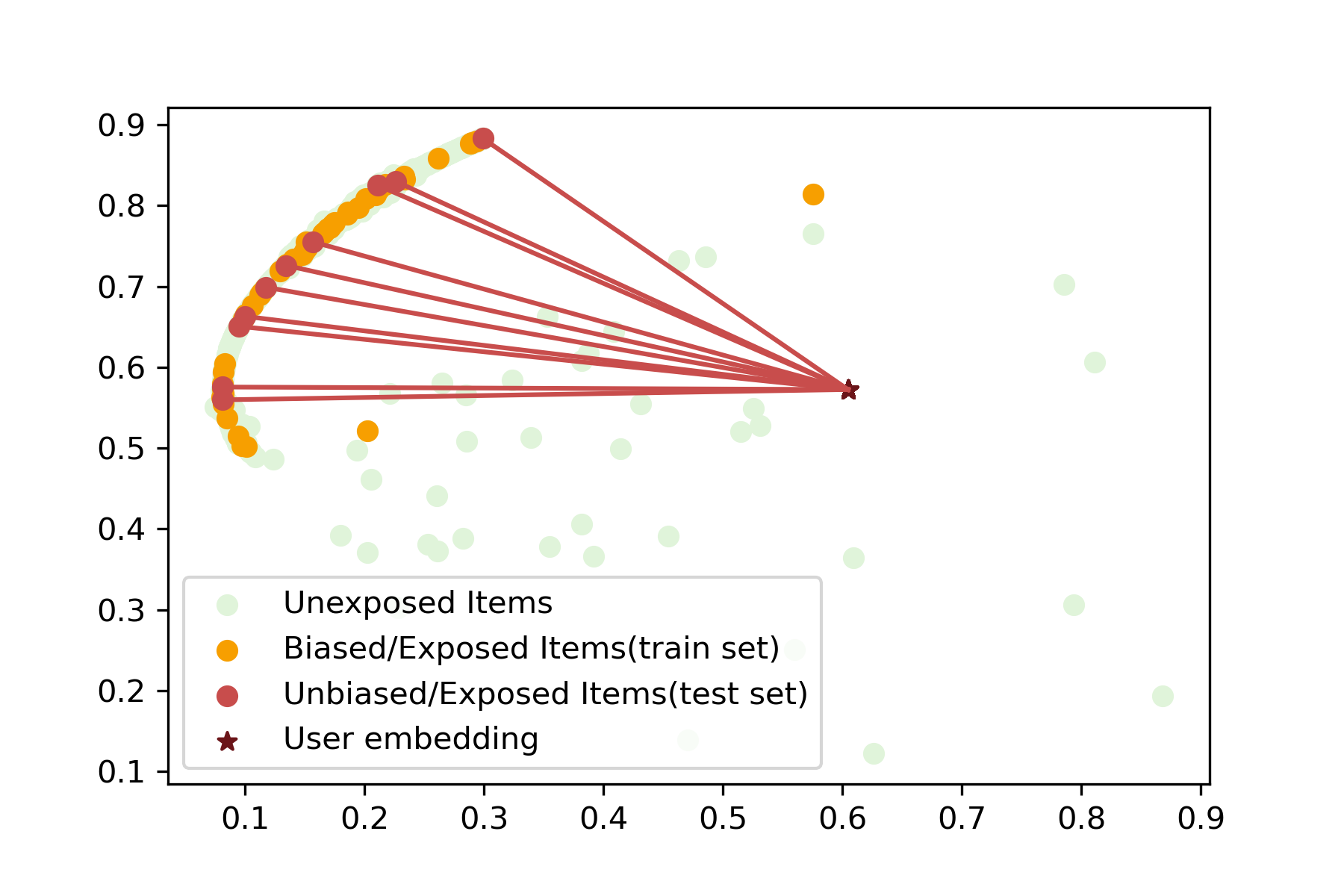}  
      \caption{CCL w/cf}
      \label{680968909872}
    \end{subfigure}
\caption{
The utilization of t-SNE visualization in the Yahoo! R3 dataset allows for the representation of both user and item. 
The green dots represent unexposed items, the orange dots represent interacted items within the training set, and the red dots represent items within the testing set. 
The sub-figures demonstrate the our method CCL with three positive sampling techniques and the method of MF\_CVIB, in the ascending order of performance on the Yahoo dataset from left to right.
}
\label{fig:representation}
\end{figure*}

\subsubsection{Visualization study}
We previously discuss the relationship between contrastive self-supervised learning (SSL) and the reduction of exposure bias.
In order to further investigate this connection, we propose three novel positive sampling techniques that are integrated in classic contrastive SSL framework.
These techniques aim to mitigate confounding bias in the representation space, where items are presented to users randomly.  
Two crucial questions that we aim to address in this section are: 
(1) whether we can manipulate exposure more uniformly in the representation space, and (2) whether this uniform exposure improves the performance of inferring user natural preferences, as tested on unbiased test data. 

To answer these questions, we conduct the t-SNE visualization \cite{van2008visualizing} to analyze the distribution of items in the user-item shared representation space for a randomly selected user.
We include items from both the training and test sets that correspond to this chosen user.
In addition to our CCL method with three sampling methods, we also utilize the MF\_CVIB method, which has demonstrated superior performance on the Yahoo dataset in Table \ref{907914275759} and \ref{364253059135}. Due to the space limitations, we only present the visualizations on the Yahoo dataset.

It is important to note that we arrange these four visualization figures in ascending order of performance on the Yahoo dataset, from left to right.
Consequently, the visualizations, shown as Figure \ref{fig:representation}, reveal that items become increasingly uniformly distributed around the randomly selected user.
As an example, taking the CCL w/cf method as illustrated in Figure \ref{680968909872}, we can observe that the majority of item dots possess a uniform distance from the user dot. 
However, there are still data points that remain scattered chaotically within the representation space.
Our analysis suggests that this may be attributed to the presence of numerous uninteracted user-item pairs, which the method is unable to fully sample.   
It should be pointed out that the location of the user dot in each sub-figure is distinct, as we randomly select a user for each figure.
Additionally, it should be acknowledged that the t-SNE algorithm alters the spatial positioning of the user representation. 
The purpose of this analysis is to examine the relative positions of items in relation to the user .

\section{Related Work}
\label{sec_relatedwork}
\subsection{Causal Inference for Recommender Systems} 
To date, a significant number of studies have linked causal inference with recommender systems, viewing recommendation as an intervention in causal inference. 
Several studies integrate the propensity score, a causality concept, into their recommendation models to achieve unbiased learning and evaluation \cite{schnabel2016recommendations, swaminathan2015self, jiang2016doubly, wang2019doubly, li2022stabilized, zhou2022cycle}. 
Other research assumes the availability of a small set of unbiased data or additional knowledge. 
For example, \cite{bonner2018causal, wang2021combating} utilize unbiased data while \cite{li2021causal, sheth2022causal} leverage network information to reduce biases and improve performance. 
However, many methods introduce extra variance or do not fully explain the data generation process.
In contrast, our approach involves analyzing the data generation and resolving exposure bias through contrastive SSL and novel sampling methods.

\subsection{Contrastive Learning For Recommendation}

Recent studies have applied contrastive SSL into recommender systems, such as \cite{zhou2021contrastive, lin2021dynamic, liu2021contrastive, yu2022self}. 
While works \cite{liu2021contrastive, yu2022self} focus on sequential recommendations with contrastive SSL, our approach concentrates on classic recommendations without any sequential information. 
Thereby, we are unable to directly adopt augmentation operators from previous studies \cite{liu2021contrastive}. 
In contrast to prior works \cite{zhou2021contrastive, lin2021dynamic}, which frame propensity score or item popularity as the negative distribution, our approach starts from the perspective of a positive sampler. 
Additionally, our best sampling method does not rely on the propensity score or item popularity, reducing computation cost.
Importantly, our contrastive SSL highlights the usefulness of counterfactual samples, offering a fresh perspective on solving causal inference for recommender systems.

\section{Conclusion}
\label{sec_conclusion}
The present study endeavors to examine the causal interpretation of recommender systems and to demonstrate the necessity of addressing the confounding bias introduced by the underlying exposure mechanism. 
To achieve this objective, we employ a contrastive self-supervised learning (SSL) framework, specifically through the utilization of inverse propensity scores and the expansion and diversification of the positive sample set. 
Furthermore, we propose three novel sampling strategies integrated into contrastive learning and introduce our method, referred to as contrastive counterfactual learning (CCL). 
Through extensive experimentation on two real-world benchmark datasets, we demonstrate the effectiveness of our approach in comparison to ten state-of-the-art baselines.

Looking ahead, our research aims to enhance standard recommendation models with counterfactual samples and to develop metrics for estimating the causal effect of recommendations \cite{sato2020unbiased}.

\clearpage

\bibliographystyle{ACM-Reference-Format}
\balance
\bibliography{sample-base}

\end{document}